\documentstyle[epsfig]{basi}

\DeclareMathAlphabet{\mathsc}{OT1}{cmr}{m}{sc}
\def\testbx{bx}%
\DeclareRobustCommand{\ion}[2]{%
\relax\ifmmode
\ifx\testbx\f@series
{\mathbf{#1\,\mathsc{#2}}}\else
{\mathrm{#1\,\mathsc{#2}}}\fi
\else\textup{#1\,{\mdseries\textsc{#2}}}%
\fi}
\def\ch{\footnotesize}
\def\HI{\ion{H}{i}~}
\def\km{km~s$^{-1}$~}
\def\deg{\hbox{$^\circ$}~}
\def\AA{Astron. Astrophys.}
\def\ApJ{Astrophys. J.}
\def\MNRAS{Mon. Not. Roy. Astron. Soc.}
\def\AJ{Astron. J.}

\begin{document}
\title[The Eridanus group]{\HI deficiency in groups : what can we learn from Eridanus}
\author[Omar]
 {A. Omar\thanks{e-mail : aomar@upso.ernet.in}\thanks{\em Present address : ARIES, Manora peak, 
 Nainital, 263 129, Uttaranchal, India} \\
	Raman Research Institute, Sadashivanagar, Bangalore, 560 080 India}

\maketitle
\label{firstpage}

\begin{abstract}

The \HI content of the Eridanus group of galaxies is studied using the GMRT
observations and the HIPASS data. A significant \HI deficiency up to a factor
of $2-3$ is observed in galaxies in the Eridanus group. The deficiency is found
to be directly correlated with the projected galaxy density and inversely
correlated with the line-of-sight radial velocity. It is suggested that the \HI
deficiency is due to tidal interactions. An important implication is that
significant evolution of galaxies can take place in a group environment. 

\end{abstract}



\section{Introduction}

\begin{figure}
\centering
\includegraphics[width=8cm, angle=0]{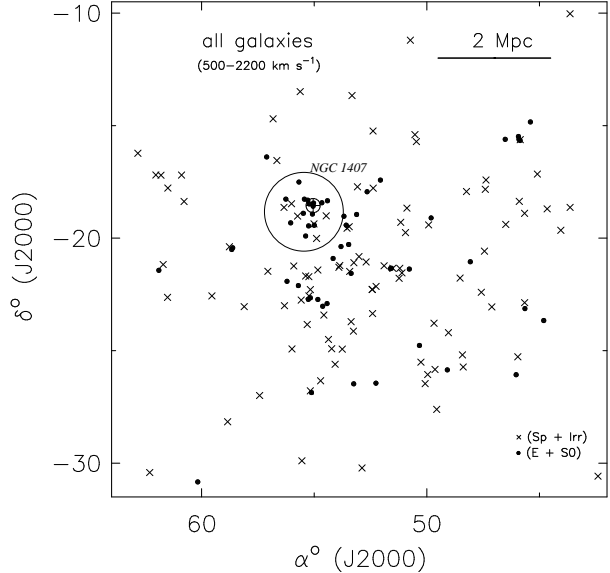}

\caption{Galaxies in the Eridanus group. The early type 
(E+S0) galaxies are marked as filled circles and late type (Sp + Irr) galaxies
are marked as crosses. NGC~1407, the brightest galaxy of the group is marked as encircled `+' sign. 
}

\label{fig:l-b}
\end{figure}

Spiral galaxies in the cores of clusters are known to be  \HI deficient
compared to their field counterparts (Davies \& Lewis 1973, Giovanelli \&
Haynes 1985, Cayatte et al. 1990, Bravo-Alfaro et al. 2000). Several
gas-removal mechanisms have been proposed to explain the \HI deficiency in
cluster galaxies. There are convincing results from both the simulations and
the observations that  ram-pressure stripping can be active in galaxies which
have crossed the high ICM (Intra Cluster Medium) density region in the cores of
clusters (Vollmer et al. 2001, van Gorkom 2003). However, it is not clear that
all \HI deficient galaxies have crossed the core. ``Galaxy harassment'' can
also affect  outer regions of  galactic disks as a result of repetitive fast
encounters of galaxies in clusters (Moore et al. 1998). There can be other
scenarios where galaxies can become gas deficient, e.g., thermal evaporation \&
viscous stripping (Cowie \& Songaila 1977, Nulsen 1982, Sarazin 1988) and
removal of the hot gas from galaxy halo (``galaxy strangulation''; Larson et
al. 1980). It turns out that no single gas-removal mechanism can explain the
global \HI deficiency in a cluster environment (e.g., Magri et al. 1988). Some
of the parameters driving these mechanisms are not known well, e.g., thermal
conductivity of the ICM, amount of hot gas in halos etc. These difficulties
have led one to speculate that cluster galaxies were perhaps \HI deficient even
before they fall in to the cluster.

Several clusters have been imaged in \HI. However, only limited \HI data exist
for large groups. For example, Ursa-Major group which has no elliptical and
only a few S0's shows no significant \HI deficiency (Verheijen 2001). Hickson
Compact Groups (HCGs) have been studied in \HI and some of them show \HI
deficiency (Verdes-Montenegro et al. 2001). However, small number of galaxies
in HCGs makes this sample statistically insignificant to make any firm
conclusion or to understand the mechanism deriving the galaxy evolution. Here,
we present  an \HI study of the large ($\sim200$ galaxies) nearby
($\sim23$~Mpc) Eridanus group of galaxies. The properties of the group are
described in the next section. Both the GMRT data and the HIPASS (\HI Parkes
All Sky Survey) data are used. The details of the GMRT observations, data
reduction and analyses are presented in Omar \& Dwarakanath (2004a). The full
description of this work is presented in the Ph.D. thesis of  Omar (2004). A
detailed discussion on the \HI content of the Eridanus group is given in Omar
\& Dwarakanath (2004b). Here, we discuss the results briefly.

\begin{table*}
\begin{center}
\caption{Comparison of four nearby galaxy groups and clusters}
\vspace{0.1in}
\label{tab:compare}
\begin{tabular}{lcccc}
\hline
\hline
\bf {Properties} & \bf{Virgo$^{a}$} & \bf{Fornax$^{b}$} & \bf{Eridanus$^{c}$} & \bf{Ursa-Major$^{d}$} \\
\hline

Distance (Mpc)			  & 21   & 20     & 23      & 21  \\
No. of E+S0                       & 71   & 23     & 36      & 9   \\
No. of S + Irr                    & 123  & 17     & 42      & 53  \\
Vel. dispersion (~\km)            & 760  & 350    & 240     & 150 \\
log X-ray luminosity (erg s$^{-1}$)& 43.5 & 41.7   & 41.4    & -- \\
Proj. galaxy density (Mpc$^{-2}$) & 50   & 70     & 8       & 3   \\

\hline 

\multicolumn{5}{p{5in}}{\ch  Notes - (a): Inner 6\deg ($\sim 2.4$ Mpc) region, (b): Inner
2\hbox{$^\circ$}.4 ($\sim 1$ Mpc) region, (c): Inner 9\deg ($\sim 3.6$ Mpc) region, (d): Inner 15\deg
($\sim 6$ Mpc) region. The table is from Omar et al. (2004). }

\end{tabular} 
\end{center} 
\end{table*}

\section{The Eridanus group}

The Eridanus group was identified as a moderate size cluster in a large scale
filamentary structure near $cz$~$\sim1500$~km~$s^{-1}$ in the Southern Sky
Redshift Survey (SSRS; da Costa et al. 1988). This filamentary structure, which
is the  most prominent in the southern sky, extends for more than 20 Mpc. The
Fornax cluster and the Dorado group of galaxies also belong to this structure. 
Willmer et al. (1989)  grouped the galaxies in the Eridanus region in to
different sub-groups. They concluded that each sub-group is a bound structure
and the entire group is also gravitationally bound with a dynamical mass
greater than $\sim10^{13}$~M$_{\odot}$.  

Most of the Eridanus galaxies are concentrated in the velocity range $cz =
1000-2200$~km~$s^{-1}$.  The distance to the group is estimated as
$\sim23\pm2$~Mpc from the surface brightness fluctuation measurements (Tonry et
al. 2001). The locations of galaxies in the Eridanus group are plotted in
Fig.~\ref{fig:l-b} where it can be seen that the galaxies are not  distributed
uniformly in space. The group appears to be made of different sub-groups. The
sub-clustering is quite prominent in the inner region. The sub-group NGC~1407
(cf. Willmer 1989; marked in Fig~\ref{fig:l-b}) has its brightest member as  an
elliptical galaxy (NGC~1407; E0) which is also the brightest in the entire
group. NGC~1407 has diffuse X-ray emission ($L_{X} = 1.6 \times
10^{41}$~erg~s$^{-1}$) surrounding it. NGC~1407 sub-group is the richest in
early types, most of them being S0's. The population mix of (E+S0) and
(Spirals+Irr) in the NGC~1407 sub-group is 70\% \& 30\% respectively. The
overall population mix in the Eridanus group is 30\% (E+S0) \& 70\% (Sp +Irr).

There is no appreciable difference in the velocities over which the early types
and the late type galaxies are distributed. This is contrary to that seen in
clusters like Virgo and Coma where spirals have much flatter velocity
distribution and E/S0's have nearly a Gaussian distribution. In
Tab.~\ref{tab:compare}, the properties of the Eridanus group are compared with
the two clusters Virgo and Fornax and one loose group Ursa-Major. It appears
that the Eridanus group forms a system which is intermediate between a loose
group (e.g., Ursa-Major) and a cluster (e.g, Virgo, Fornax). It is interesting
to note that the population mix in the NGC~1407 sub-group is similar to that
seen in evolved clusters.

\begin{figure}
\centering
\includegraphics[width=13cm, angle=0]{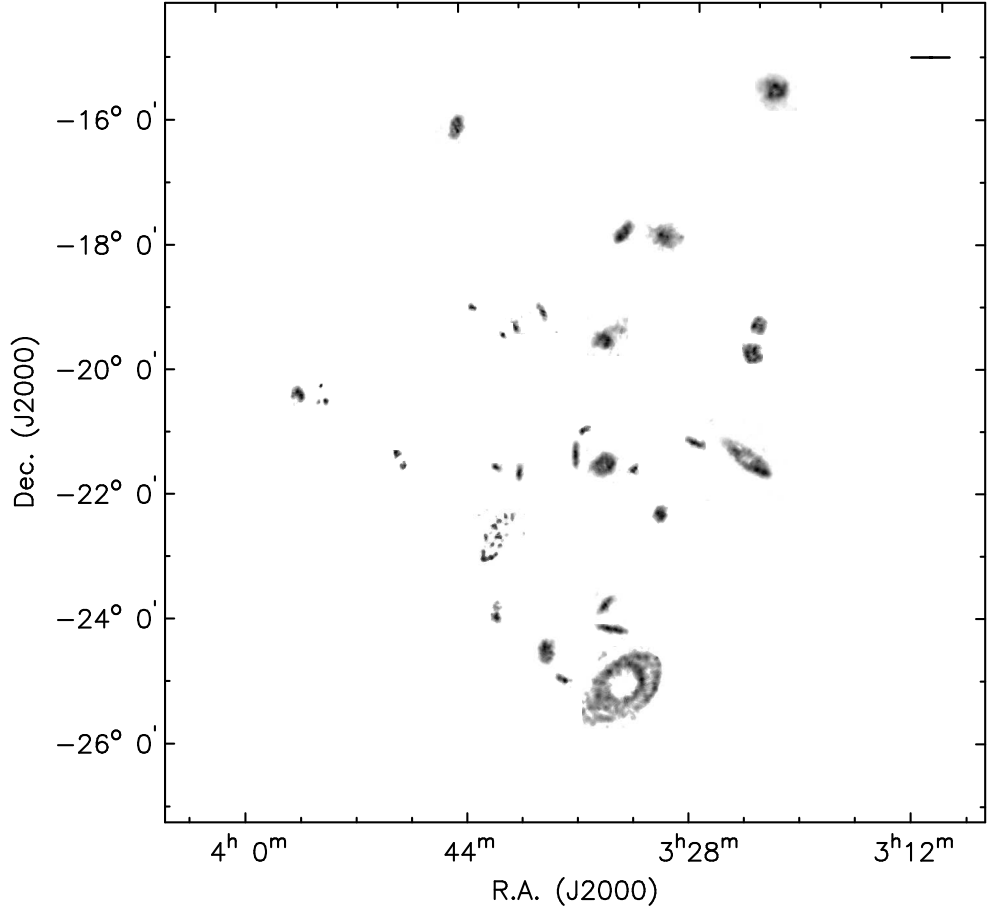}

\caption{A collage of the total \HI maps of the Eridanus galaxies from GMRT. The individual galaxies are
enlarged by $\sim10$ times their original size. The actual positions of some galaxies are displaced
to avoid overlap. A bar on the upper right hand corner indicates a scale of 20~kpc for the enlarged
sizes of the galaxies, otherwise, $1\deg$ corresponds to a linear dimension of $\sim400$~kpc. The
maps are sensitive down to an $5\sigma$ column density of $N_{\HI} = 10^{20}$~cm$^{-2}$.}

\label{fig:grpview}
\end{figure}

\section{Results}

\begin{figure}
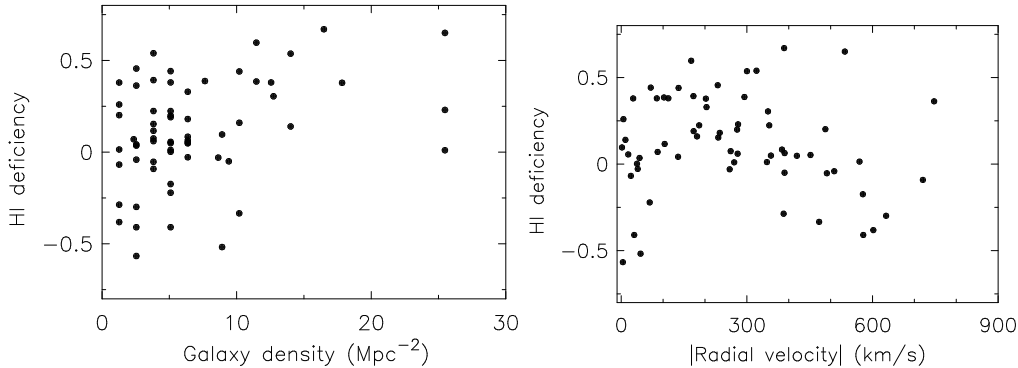


\resizebox{6.8cm}{!}{\includegraphics[angle=0]{dens_defi.epsi}}
\resizebox{6.5cm}{!}{\includegraphics[angle=0]{velo_defi.epsi}}

\caption{ (left): \HI deficiency plotted against the projected galaxy density. (right): \HI
deficiency plotted against the line-of-sight radial velocity. }

\label{fig:def}
\end{figure}

The galaxies in the Eridanus group were observed in the \HI 21cm-line with the
GMRT (Omar \& Dwarakanath 2004a). A collage of the GMRT \HI images of the
Eridanus galaxies is shown in Fig.~\ref{fig:grpview}. The data from the HIPASS
(Meyer et al. 2004) were also used. The final sample consisted of a total of 63
\HI detected galaxies.  The deficiency parameter (cf. Haynes \& Giovanelli
1984, $hereafter$ HG84) estimated using  $log (M_{\HI}$/$D_{opt}^{2})$ ratio
was used to obtain the \HI deficiency in the Eridanus galaxies. The ratio $log
(M_{\HI}$/$D_{opt}^{2})$ for each Eridanus galaxy is compared with the mean
value of $log (M_{\HI}$/$D_{opt}^{2})$ obtained by HG84 for the isolated
galaxies of similar types. A significant positive difference in the two ratios
indicates the \HI deficiency (def. =  $<log(M_{\HI}/D_{opt}^{2})>_{field} -
~log(M_{\HI}/D_{opt}^{2})$). The \HI deficiency up to a factor of $2-3$
($\sim0.5$ in log units in Fig.~\ref{fig:def}) is seen in the Eridanus
galaxies. In Fig.~\ref{fig:def}, \HI deficiency is plotted against the local
projected galaxy density and the line-of-sight radial velocities of galaxies in
the group. The projected galaxy density is estimated within a circular region
of radius 0.5~Mpc. It can be seen that in  higher galaxy density
($>10$~Mpc$^{-2}$) regions, majority of galaxies are \HI deficient while in
lower galaxy density regions both normal and deficient galaxies are present.
The \HI deficiency appears to be increasing with decreasing line-of-sight
radial velocity.

\section{Discussion}

\begin{figure}
\centering
\
\includegraphics[width=7cm, angle=0]{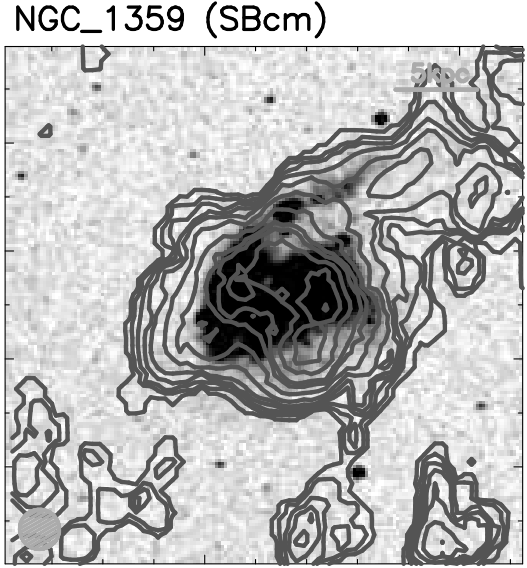}

\caption{GMRT \HI contour image overlaid upon the optical DSS image of NGC~1359
which shows both gaseous  and optical tidal features. The contours start at
$N_{HI} = 10^{20}$~cm$^{-2}$.} 

\label{fig:collage} 

\end{figure}

The ram-pressure in the Eridanus group is one to two orders of magnitude lower
compared to that in clusters (Omar \$ Dwarakanath 2004b). Therefore,
ram-pressure stripping is of a little importance in the Eridanus galaxies. The
trends in Fig.~\ref{fig:def} are suggestive of the \HI deficiency being due to
tidal interactions. Since galaxies in higher galaxy density regions will have
higher probability of tidal encounters, it explains the increasing trend of
deficiency with increasing galaxy density. The inverse correlation between the
deficiency and the radial velocity can be qualitatively understood under the
framework of tidal interactions. The perturbation due to tidal interactions
will be maximum for slow encounters. In the Eridanus group where the
distribution of galaxies is peaked near the mean velocity of the group (Omar
2004) and falls off nearly as a  Gaussian at higher relative velocities,
galaxies with velocities near the mean velocity of the group will have a higher
probability of interacting with a companion with a lower velocity difference. 
Both gas and stars will be removed from galaxies as a result of tidal forces.
It was noticed that galaxies with larger optical diameters are predominantly in
the lower galaxy density regions (Omar \& Dwarakanath 2004b).  The Eridanus
galaxies often show signatures of tidal interactions, e.g., shrunken \HI disks,
warps, asymmetric \HI disk, \HI rings, tidal tails, extraplannar gas,
kinematical lopsidedness etc. The GMRT \HI image of one such galaxy having both
the gaseous and the optical tidal tail is shown in Fig.~\ref{fig:collage}. More
images can be found in Omar \& Dwarakanath (2004b).

If clusters are built via mergers of small groups, the \HI deficiency in the
Eridanus group indicates that not all \HI deficiency in cluster galaxies
originates in the cluster environment. Alternatively, a significant fraction of
the \HI deficiency in cluster galaxies could have originated in the group
environment. The implications are discussed in detail in Omar \& Dwarakanath
(2004b).

\section*{Acknowledgments}

This paper is based on the talk presented in the meeting at NCRA, Pune held
during $22-23$ March, 2004 on the occasion of 75$^{th}$ birthday of Prof.
Govind Swarup. The work presented here is based on the Ph.D. thesis of A. Omar
(2004). We thank the staff of the GMRT who made observations possible. This
research has made use of the HI Parkes All Sky Survey (HIPASS) data.

\label{lastpage}


\begin{thebibliography}{10}

\bibitem{bravo} Bravo-Alfaro, H., Cayatte, V., van Gorkom, J. H., \& Balkowski, C. 2000, {\it
\AJ}, {\bf 119}, 580
\bibitem{cay} Cayatte, V., van Gorkom, J. H., Balkowski, C., \& Kotanyi, C. 1990,
{\it \AJ}, {\bf 100}, 604
\bibitem{Cowie} Cowie, L. L., \& Songaila, A. 1977, {\it Nature}, {\bf 266}, 501
\bibitem{daCos88} da Costa, L.N., Pellegrini, P.S., Sargent, W.L. et al. 1988, {\it \ApJ}, {\bf 327}, 544
\bibitem{dav} Davies, R. D. \& Lewis, B. M. 1973, {\it \MNRAS}, {\bf 165}, 231
\bibitem{GH85} Giovanelli, R., \& Haynes, M. P. 1985, {\it \ApJ}, {\bf 292}, 404
\bibitem{HG84} Haynes, M. P., \& Giovanelli, R.  1984, {\it \AJ}, {\bf 89}, 758
\bibitem{larson} Larson R.B. Tinsley, B.M., \& Calswell, C.N. 1980,{\it \ApJ}, {\bf 237}, 692
\bibitem{magri} Magri, C., Haynes, M. P., Forman, W., Jones, C., \& Giovanelli, R. 1988,
{\it \ApJ}, {\bf 333}, 136
\bibitem{hipass} Meyer, M.J., Zwaan, M.A., Webster, R.L. et al. 2004, {\it \MNRAS}, {\bf 350}, 1195
\bibitem{moore} Moore, B., Lake, G., \& Katz, N. 1998, {\it \ApJ}, {\bf 495}, 139
\bibitem{Nulsen} Nulsen, P.E.J. 1982, {\it \MNRAS}, {\bf 198}, 1007
\bibitem{omar} Omar, A. 2004, Ph.D. Thesis, Jawaharlal Nehru University, New Delhi.
\bibitem{omar} Omar, A. \& Dwarakanth K. S. 2004a, {\it J. Astroph. Astron.}, submitted
\bibitem{omar} Omar A. \& Dwarakanth K. S. 2004b, {\it J. Astroph. Astron.}, submitted
\bibitem{Sarazin} Sarazin, C.L., 1998, {\em In X-ray emission from clusters of galaxies, Cambridge Astroph.
Series, University Publications.}
\bibitem{ton01} Tonry, J.L. et al. 2001, {\it \ApJ}, {\bf 546}, 681
\bibitem{vanG03} van Gorkom, J.H. 2003, {\it In clusters of galaxies: Probes of cosmological
structure and galaxy formation, ed. Mulchaey, J.S et al., (Carnegie Obs. Astroph. Ser.)}, {\bf 3}
\bibitem{HCG} Verdes-Montenegro, L., Yun, M. S., Williams, B. A., Huchtmeier, W. K., Del
Olmo, A., \& Perea, J. 2001, {\it \AA}, {\bf 377}, 812
\bibitem{ver} Verheijen, M.A.W, 2001, {\em In gas and galaxy evolution, ed. Hibbard, J.E. et al., ASP
conf. series}, {\bf 240}, 573
\bibitem{vol01} Vollmer, B., Cayatte, V., Balkowski, C., \& Duschl, W. J. 2001, {\it \ApJ},
{\bf 561}, 708 
\bibitem{wil89} Willmer, C.N.A., Focardi, P., da Costa, L.N., \& Pellegrini, P.S. 1989, {\it
\AJ}, {\bf 98}, 1531

\end{thebibliography}
\end{document}